\documentclass[12pt]{article}
\usepackage{epsfig,pstricks}

\def\f{\varphi}
\def\l{\lambda}
\def\G{\Gamma}
\def\gf{\gamma_\f}
\def\fb{\bar\f}

\def\p{\partial}

\def\loopd#1{\hbox{$\raisebox{0.8pt}{$\bigcirc$}\kern-9pt{#1}\kern+5pt$}}
\def\o{\over}
\def\be{\begin{eqnarray}}
\def\ee{\end{eqnarray}}
\def\ds{\displaystyle}
\def\ba{\begin{array}}
\def\ea{\end{array}}
\def\b{\beta}

\begin{document}
\setcounter{page}{0}
\thispagestyle{empty}
\begin{flushright}
May 6, 2004
\end{flushright}
\vskip 22 truept
\begin{center}
{\LARGE Dimensional Reduction, Hard Thermal Loops and the \\[2mm]
Renormalization Group{\Large\footnote{This work was
supported by Conacyt grant 32399--E.}}}
\vskip 0.4truein
{\bf C.R. Stephens}\footnote{e--mail: 
stephens@nuclecu.unam.mx}$^{,a}${\bf,
Axel Weber}\footnote{Supported by CIC--UMSNH, e--mail:
axel@itzel.ifm.umich.mx}$^{,b}${\bf,
Peter O. Hess}\footnote{e--mail: 
hess@nuclecu.unam.mx}$^{,a}${\bf, 
Francisco Astorga}\footnote{Supported by CIC--UMSNH, e--mail:
astorga@ifm1.ifm.umich.mx}$^{,b}$
\vskip 0.25truein
${}^a\,$Instituto de Ciencias Nucleares, UNAM, \\
Circuito Exterior C.U., A. Postal 70-543, \\
04510 M\'exico D.F., Mexico. \\[5mm]
${}^b\,$Instituto de F\'{\i}sica y Matem\'aticas, Universidad 
Michoacana de San Nicol\'as de Hidalgo, \\
Edificio C--3 Cd.\ Universitaria, A. Postal 2--82, \\
58040 Morelia, Michoac\'an, Mexico
\end{center}
\vskip 0.6truein
{\bf Abstract:}\
We study the realization of dimensional reduction and the validity of
the hard thermal loop expansion for $\l\f^4$ theory at finite
temperature, using an environmentally friendly finite--temperature 
renormalization group with a fiducial temperature as flow parameter.
The one--loop renormalization group allows for a consistent description of 
the system at low and high temperatures, and in particular of the phase
transition. The main results are that dimensional reduction applies, apart
from a range of temperatures around the phase transition, at high 
temperatures (compared to the zero temperature mass) only for sufficiently
small coupling constants, while the HTL expansion is valid below (and
rather far from) the phase transition, and, again, at high temperatures
only in the case of sufficiently small coupling constants. We emphasize
that close to the critical temperature, physics is completely dominated
by thermal fluctuations that are not resummed in the hard thermal loop
approach and where universal quantities are independent of the 
parameters of the fundamental four--dimensional theory.
\vfill\eject

\section{Introduction}

Finite temperature field theory is of great importance in many areas
of physics.  In the realm of high energy physics there are two main
areas: phase transitions in the early universe and heavy ion
scattering, where the phase structure of QCD may potentially be
studied. One of the most salient features of finite temperature
physics is that the effective degrees of freedom  may be quite
different at high temperature than at low temperature.  Not the least
of these changes is the potential change in effective dimensionality
of a system at ``high'' temperature. We put ``high'' in quotes to
emphasize that what is meant by high is a quite subtle issue, as are
the  nature and circumstances of dimensional reduction. Naturally,
the question arises as to ``high'' relative to what?

Many of the standard tools of field theory have been applied to finite
temperature field theory, ranging from straightforward perturbation
theory and other expansions, such as $1/N$, to various forms of the
renormalization group (RG). The fact that understanding the high
temperature regime is not trivial can be illustrated by some of the
erroneous results that have been derived, such as that there is
asymptotic freedom at  high temperature in QCD, or that
$\lambda\varphi^4$ at finite temperature exhibits a first order
transition in the absence of an ordering field.  

As is well known, at temperatures $T$ of the order of $M_0/g$ or higher, 
the perturbative expansion breaks down. Here $M_0 \equiv M(T=0)$ is the
zero--temperature mass and $g$ is a generic zero--temperature coupling
constant, for example $g = \sqrt{\lambda_0}$ (with $\lambda_0 \equiv
\lambda (T=0)$) in a scalar $\lambda
\varphi^4$ theory. The reason for the failure of straightforward perturbation 
theory is the appearance of a thermal mass $\sim g T$ which acts as an IR 
cutoff for the momentum integrals in Feynman diagrams and generates inverse 
powers of $g$ in the perturbative expansion. The physical origin of these
infrared problems is that at high temperature the effective degrees of
freedom are quite different to those at zero temperature.

Through a careful analysis of perturbation theory it can be shown that
these problems can be cured by an appropriate resummation of the most
infrared divergent graphs at high temperatures, the so--called hard thermal
loops (HTL) \cite{htl}. In the case of $\lambda \varphi^4$ theory, it is
sufficient to replace the zero--temperature mass $M_0$ in the propagators
by the temperature--dependent mass $M(T)$, calculated consistently order 
by order, while the coupling constants may be treated perturbatively once 
the mass corrections have been resummed (see \cite{parwani} for the 
explicit results to two loops in the resummed theory).

However, the HTL approximation is invalid near a second order or weakly 
first order transition where the temperature--dependent mass $M(T)$
becomes small and the resulting IR divergencies are of a qualitatively 
different nature: the infrared fluctuations dominate in the same 
sense as in critical phenomena, where they invalidate a perturbative treatment 
of all physical quantities. For instance, the existence of another RG
fixed point invalidates a perturbative expansion around the Gaussian
fixed point which in this case is infrared unstable. A key property of
the physics near the phase transition is universality --- universal 
quantities near the phase transition are completely independent of the zero 
temperature parameters of the theory. This is clearly not the case 
in the regime which can be described by the HTL resummation, where the zero 
temperature coupling is affected only slightly by thermal fluctuations.

Henceforth, we will refer to the regime near a phase transition, with
$M(T) \ll M_0, g T$, as the universal high temperature (UHT) regime, in
contradistinction to the non--universal high temperature (NHT) regime where 
$T \sim M_0/g$ or larger but $M(T)$ is {\em not}\ small in the above 
sense. A typical example of a relevant UHT situation would be electroweak 
theory in the vicinity of the electroweak phase transition (assuming
that the phase transition really is, at most weakly of first order), 
while an example of a NHT situation would be the electric sector of
QCD at high temperatures. One of the main  theoretical tools to access 
the NHT regime in QCD has been precisely the HTL expansion \cite{htl}.
The question then  is: under what physical circumstances is one
regime reached versus the other as a function of temperature and to
what extent the two regimes may be treated on an equal footing in a
given calculational scheme? 

The answer to the question of which regime is reached by
increasing temperature depends on whether or not a state of broken symmetry
exists at $T=0$. If the state at $T=0$ is symmetric then the mass 
$M(T)$ tends to monotonically increase as a function of temperature. 
At temperatures $g T \sim M_0$, the perturbative expansion around 
the zero temperature theory breaks down in a non--universal way (NHT 
regime). Eventually, a regime is reached where $g T \gg M_0$ and the 
physics becomes independent of the value of the zero--temperature mass. 

If, on the other hand, there is a broken symmetry at $T=0$, characteristically 
the mass will diminish until a phase transition is reached. If the phase
transition is second order or weakly first order then near the phase
transition $M(T) \ll M_0$ and perturbation theory will again break
down, but now in a universal way (UHT regime). However, there may also 
exist a NHT regime below the transition, where $g T\sim M_0$ but $M(T)$ 
is not (yet) very small relative to $M_0$, hence in this regime 
perturbation theory will break down in a similar manner to the case 
without symmetry breaking. 
Above the transition temperature, in the symmetric phase, the mass will tend 
to increase thus entering (again) a NHT regime. For $M(T) \gg M_0$, 
the physics is expected to be insensitive even to the symmetry of the zero
temperature state.

Realizing that the regime of small mass $M(T)$ is dominated by universal
fluctuations attempts have been made to describe it using techniques
gleaned from critical phenomena. An example of such an approach can be 
found in \cite{gleiser}. An alternative approach has been to describe
the regime numerically \cite{shaposhnikov}. Such attempts have all been 
based on an assumption --- that a dimensional reduction takes place and 
that the low energy limit of the theory may be described by an effective three
dimensional theory. In the case of \cite{shaposhnikov} the parameters of 
this effective theory are calculated in terms of the
parameters of the four dimensional theory and the temperature. However,
the matching of the four dimensional to the three dimensional theory is
done via a perturbative analysis analogous to the HTL expansion, so the obvious
question arises: is a dimensionally reduced NHT regime reached,
in which case a perturbative matching can be performed, before 
having to fully take into account all important universal infrared
corrections? (Note that in the case of a non--abelian gauge theory a 
perturbative treatment is only possible below the critical temperature, 
i.e., in the phase of broken symmetry.) These questions in fact, and answers 
to them, are the principle subject of this paper. 

As far as we are aware the only techniques capable of unambiguously 
answering these
questions have their origin in the RG and, in particular, must be
``environmentally friendly'' \cite{oconnstephens} RGs in that they
must be capable of describing  quantitatively the change in effective
degrees of freedom in the system as a  function of ``environment'' ---
in this case temperature. It naturally follows that such RGs must
necessarily be temperature dependent. However, temperature dependence
is not sufficient to guarantee success as various temperature
dependent RGs have been used without success (see for example
\cite{fujimoto}) to access the UHT regime. The important
requirement is that the renormalized finite temperature parameters
give an adequate description of the physics at any temperature.

Such environmentally friendly RGs have been used to successfully
investigate both the NHT and UHT regimes within one
calculational scheme, both running the thermal mass and the
temperature. For instance, in \cite{vaneijck} the critical temperature
and amplitude ratios were calculated for a $\lambda\varphi^4$ theory, thus
showing that the critical temperature, albeit a non--universal quantity,
could be calculated directly using RG techniques. The emphasis in this 
paper is on using environmentally friendly renormalization to understand when
and where a dimensionally reduced description of the system is
appropriate, what is its nature --- universal or non--universal, and
where and when important calculational techniques such as the HTL
expansion are valid. We take, again, a scalar theory as a testing
ground. 

The structure of the paper is as follows. In Section 2 we briefly
present our finite temperature renormalization prescription and the
resultant flow  equations. In Section 3 we present the solutions of the
flow equations in the NHT and UHT regimes, and consider the 
validity of dimensional reduction and the HTL expansion and finally in 
section 4 we draw some conclusions.

\section{Finite--temperature renormalization}

We define our model via the bare Euclidean action at temperature
$T=1/\beta$
\be
S[\f_{B}]=\int_0^\b dt\int
d^{d-1}x\left[{1\o2}(\nabla\f_{B})^2+{1\o2}M^2_{B}\f_{B}^2+
{\l_{B}\o4!}\f_{B}^{4}\right]\label{action}
\ee
In $d=4$ ultraviolet divergences require renormalization
of the parameters $\l_{B}$ and $M^2_{B}$, and also a wave--function
renormalization. In $d=3$ the theory is super--renormalizable
with respect to ultraviolet divergences. However, the massless
theory can give rise to infra--red divergences. Hence, we will 
define renormalized parameters via the normalization conditions
\be
{\p\o\p p^2}\left.\G^{(2)}_t(p,\fb(\tau),M(\tau),\l(\tau),T=\tau)
\right|_{p=0}&=&1,  \label{wavefunction}\\
\G^{(2)}(p=0,\fb(\tau),M(\tau),\l(\tau),T=\tau)&=&M^2(\tau),
\label{massCnd}\\[2mm]
\G^{(4)}_t(p=0,\fb(\tau),M(\tau),\l(\tau),T=\tau)&=&\l(\tau).
\label{normcnds}
\ee
The renormalization is at an arbitrary temperature $\tau$, hence 
the renormalized parameters are temperature dependent and therefore,
in principle, ``environmentally friendly''. Note that the normalization
conditions (\ref{wavefunction}) and (\ref{normcnds}) are applied to the
``transverse'' vertex functions $\G^{(2)}_t$ and $\G^{(4)}_t$ which are 
defined via the conditions
\be
\G^{(1)}=\G^{(2)}_t\fb=J,\label{eqnofstate}
\ee
which is just the equation of state for the model in the presence of
an external current $J$, $\fb$ is the renormalized field expectation value,
and  
\be
\G^{(2)}=\G^{(2)}_t+{\G^{(4)}_t\o3}\fb^2.\label{dfnofgft}
\ee
Although the transverse functions are motivated by considerations
of the $O(N)$ model, where (\ref{dfnofgft}) is a Ward identity for instance, 
they are equally valid in the analytically continued limit $N\rightarrow 1$
and lead to significant advantages in the renormalization when compared to 
normalization conditions on the ``longitudinal'' vertex funtions.
More details can be found in \cite{vaneijck}. 
Note that the physical mass $m(T)$ at the normalization point is given by
\be
m^2(\tau)=
{M^2(\tau)\o 1+{\fb^2\o3}{\p\o\p p^2}\left.\G^{(4)}_t(\tau)\right|_{p=0}}
\ee
and coincides with $M(\tau)$ when $\fb=0$ but not otherwise.

The RG flow equations are
\be
\tau{d\ln Z_\f(\tau)\o d\tau}=\gf,\qquad
\tau{dM^2(\tau)\o d\tau}=\beta_M,\qquad
\tau{d\l(\tau)\o d\tau}=\beta_\lambda.
\ee
and take different functional forms in the broken and symmetric phases. 
In a perturbative treatment of these flow equations we treat $\l$ in 
perturbation theory, but for $M$ in each diagram we eliminate $M_B^2$ in
favour of $M^2$ with the aid of condition (\ref{massCnd}).
This eliminates all diagrams which contain a tadpole as a sub--diagram.
Subsequently, we differentiate with respect to $\tau$ and solve
for the flow functions, expanding them to the loop order we are working.
It has been verified to two--loops that the resulting flow equations are free
of UV and IR divergences.

The beta functions at one loop are 
\be
\gf & = & 0, \\[2mm]
\beta_M & = &\left\{\ba{ll} \ds {\l\o 2} \tau{\p\loopd1\o\p\tau},& 
\tau>T_c\\[4mm]
\ds
-\l \left(\tau{\p\loopd1\o\p\tau}+{3\o2}M^2\tau{\p\loopd2\o\p\tau}\right),
&\tau<T_c, \ea\right.\label{massflow}\\[2mm]
\beta_\lambda & = &-{3\o2}\l^2\tau{d\o d\tau}\loopd2 .\label{couplingflow}
\ee
where the symbol $\loopd{k}$ stands for the one--loop diagram with $k$ 
propagators, without vertex factors, at zero external momentum.
It can be obtained from the following basic diagram in $d$ dimensions
(in the present case, we will consider the limit $d \to 4$)
\be
\bigcirc&=&
\tau\sum_{n=-\infty}^\infty\int{d^{d-1}k\o(2\pi)^{d-1}}
\ln(k^2+(2\pi n\tau)^2+M^2)\nonumber\\
&=&-{\G(-{d\o2})M^d\o{(4\pi)}^{d/2}}-{2\tau^d\o{(4\pi)}^{(d-1)/2}\G({d+1\o2})}
\int_0^\infty dq{q^d\o\sqrt{q^2+{z^2}}}{1\o e^{\sqrt{q^2+{z^2}}}-1},
\label{loopdef}
\ee
where $z=M/\tau$, by differentiations with respect to $M^2$.
The first derivative gives
\be
{d\o dM^2}\bigcirc=\loopd1,
\ee
whereas for $k\geq1$ we have the general rule that the derivative
with respect to $M^2$ of the loop with $k$ propagators
gives $(-k)$ times the loop with $k+1$ propagators.
The integration of equations (\ref{massflow}--\ref{couplingflow})
was used in \cite{vaneijck} to investigate the critical regime, in 
particular showing how the critical temperature could be derived using 
RG methods and also calculating critical amplitudes associated with the 
transition.

Equation (\ref{couplingflow}) is easily solved, giving 
\be
\l^{-1}(\tau)=\l^{-1}(\tau_0)
+{3\o2}\left[\loopd2(M(\tau),\tau)-\loopd2(M(\tau_0),\tau_0)\right],
\label{lambdaSolution}
\ee
where to ensure that the same initial conditions are imposed on both sides of 
$T_c$ one may use the requirement that the bare coupling $\l_B$ is the same 
in both phases to find
\be
\l^{-1}_+(\tau_+)=\l^{-1}_-(\tau_-)+{3\o2}
\left[\loopd2(M_+,\tau_+)-\loopd2(M_-,\tau_-)\right].
\label{match}
\ee
After solving the flow equations and setting the arbitrary RG temperature 
$\tau$ equal to the physical temperature $T$ the parameters 
$M(T)$ and $\lambda(T)$ describe the behaviour
of the vertex functions $\G^{(2)}$ and $\G^{(4)}_t$ at zero momentum.

Here, we have used temperature as an RG flow parameter. There is also benefit
in using a fiducial value of the finite temperature mass as flow parameter.
In this case defining the floating coupling $h=4 \l M^2 \loopd{3}$, chosen so 
that the quadratic term of the beta function has unit coefficient, one finds
\be
M{\p h\o\p M}=-(4-d_{\scriptstyle\rm eff})h + h^2
\ee
where $d_{\scriptstyle\rm eff}$ defines the effective dimension of the 
system \cite{oconnstephens}. By changing variable $M\rightarrow z=M/\tau$ we 
have $d_{\scriptstyle\rm eff}(z)$. In the limit $z\rightarrow \infty$ one 
finds that $d_{\scriptstyle\rm eff}\rightarrow 4$, with $h$ being proportional 
to the zero--temperature coupling, while in the limit $z\rightarrow 0$
one finds that $d_{\scriptstyle\rm eff}\rightarrow 3$. Thus, we see that
$d_{\scriptstyle\rm eff}$ can usefully be employed as a measure of the 
effective dimension of the system. Note that this definition of effective
dimension is universal. In other words universal quantities, such as 
critical exponents, will interpolate between those associated with different
dimensionalities. $d_{\scriptstyle\rm eff}$ will not a priori be 
suitable as a measure of non--universal dimensional reduction as in that 
case all ``universal'' quantities are mean--field like due to the 
suppression of infrared fluctuations.

\section{Results and Comparison with other Methods} 

In this section we will present results obtained by integration of the 
differential equations (\ref{massflow}--\ref{couplingflow}) for the 
temperature dependence of the mass and coupling constant. We considered 
both the case when the zero temperature theory exhibited a broken symmetry 
and when it did not. Additionally, in both these cases we considered 
different initial conditions for the zero temperature coupling.

As mentioned, our principle concerns are to examine the phenomenon of 
dimensional reduction and also the regime of validity of the HTL 
approximation. Normally dimensional reduction is something that is assumed 
based on simple dimensional arguments. For example, Fourier transforming 
in the imaginary time formalism one finds an infinite tower of masses
$M_0^2+4\pi^2T^2n^2$. One argues that in the infrared only the lowest mode,
$n=0$, is relevant, the higher mass modes decoupling in the infrared
for sufficiently high temperature $T$ ($T\gg M_0$), thus leading 
to an effective three dimensional theory. The problem with this is, as we have 
already discussed, there are two quite distinct regimes where this can take 
place --- the UHT and NHT regimes --- and, also as we have emphasized, these
are universal and non--universal respectively. Hence, in the UHT
regime dimensional reduction will be a universal phenomenon, 
independent of the zero temperature theory, whereas in the NHT
limit its existence will depend on the zero temperature parameters.
In the UHT regime, due to the existence of universal fluctuations,
one would expect to see three dimensional critical behaviour with
associated three--dimensional critical exponents. This is indeed what
is found. However, the NHT regime is not fluctuation
dominated due to the suppression of infrared fluctuations by the large
thermal mass. In this case one expects the behaviour to be ``mean field'' like.
Once again, this is what is observed.

On a qualitative level, in the imaginary--time formalism, $\beta = 1/T$ 
is the extension of space--time in the temporal direction. As long as the 
correlation length $\xi = 1/M$ is much smaller than $\beta$, or equivalently 
$T \ll M(T)$, the system does not ``feel'' the restriction in the temporal 
direction, while for $\xi \gg \beta$ or $T \gg M(T)$, the field configuration 
can be considered as approximately constant in this direction as long as we 
consider modes with four--momenta $p < T$, and the system reduces to
an effectively three--dimensional one. This criterion replaces the naive
one according to which dimensional reduction occurs whenever  $T \gg M_0$
from a consideration of the bare propagator in the Matsubara sum
representation. The difference consists precisely in considering the
renormalized or ``resummed'' mass $M(T)$ instead of the zero--temperature
mass $M_0$. Thus, we may fix a criterion for dimensional reduction that
there is dimensional reduction when $z(T) = M(T)/T < z_0$, where
$z_0$ is some suitably chosen number. Typically, we will choose $z_0=0.3$ 
so that dimensional reduction is considered to be valid when the system 
``size'' is 0.3 times the correlation length. In the plots of $M(T)$ vs.\
$T$ below, a fixed value of $z_0$ corresponds to a straight line through
the origin with slope $z_0$. For the part of the curve $M(T)$ below this
line we then consider a dimensional reduction to have occurred while above it 
not.

\begin{figure}
\begin{center}
\pspicture(10,6.3)
\rput[lb]{0}(0,-0.5){\includegraphics[width=10cm]{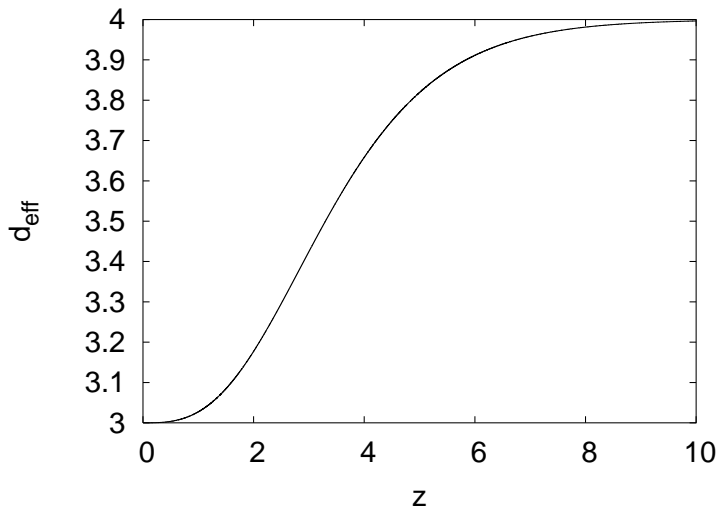}}
\endpspicture
\end{center}
\caption{The effective dimension defined in Eq.\ (\ref{defdeff}), as a
function of $z = M/T$. \label{deff}}
\end{figure}

Of course, the crossover from four to three dimensions is smooth. The 
effective dimension introduced in the previous section takes this crossover 
into account in a natural fashion.    
For instance, using the effective dimension, if we consider the regime 
where $z(T) = 0.3$ we may ask to what
extent there is a dimensional reduction, i.e.\ to what extent the system
appears to be three dimensional. Explicitly,  
\be
d_\mathrm{eff} &=& 4 + \frac{\ds \rule[-1mm]{0mm}{8mm} \frac{\p^2 \loopd2}
{(\p \ln M)^2}}{\ds \rule[-1mm]{0mm}{8mm} \frac{\p \loopd2}{\p \ln M}} 
\;\;=\;\; 6 - 6 M^2 \frac{\ds \rule[-1mm]{0mm}{5mm} \loopd4}
{\ds \rule[-1mm]{0mm}{5mm} \loopd3} \:. \label{defdeff}
\ee
which turns out only to 
depend on $z$. We note that it is essential to keep the logarithmic term
in the expressions for the one--loop diagrams because it represents the
zero temperature contribution. We have plotted $d_\mathrm{eff}$ as a function 
of $z$ in Fig.\ \ref{deff}. 
The limit $z \to \infty$
corresponds to $\tau \to 0$ (for a finite mass $M_0$), where physics is
entirely four--dimensional, whereas $z \to 0$ holds at the phase transition
where $M(\tau) \to 0$ at finite $\tau$, and a universal dimensional reduction 
to three dimensions is fully realized. We can use $d_\mathrm{eff}$ to justify 
our choice of $z_0 = 0.3$ noting that for this value of $z$ the effective 
dimension is equal to three with good precision (less than 3.01).  
Further evidence for  what values of $z_0$ lead to dimensional reduction 
can be gleaned from considerations of any universal physical quantity. In 
particular in \cite{oconnstephens,freirespecific} effective critical exponents 
that interpolate between two different dimensions were considered in the 
context of a finite size system in a film geometry of thickness $L$ with 
periodic boundary conditions (mathematically equivalent to the
present case). It was found that the crossover region is broader crossing
over from four to three dimensions rather than three to two. Also, it was found
that the crossover took place at smaller values of $z_0$ for four to three
crossover than for three to two. Taken together these two points led to the 
observation that three--dimensional critical behavior was not realized until 
$T\sim 100M(T)$ at least. In light of this evidence our criterion of $z_0=0.3$
is quite liberal as, although in this case there were 
deviations from four--dimensional behavior, one could not argue that the 
behavior was effectively three--dimensional.

Our other principal concern is the validity of the HTL expansion. 
The HTL expansion devises a systematic scheme of partial resummation of
the standard perturbation series in the case of finite temperature, thus
recovering an expansion in powers of the coupling constant (actually, in
powers of $\sqrt{\l_0}$ in the present case), while a naive perturbative
expansion runs into problems for ``soft'' external momenta $p < \sqrt{\l_0}
\, T$ for temperatures of the order $M_0/\sqrt{\l_0}$ or higher,
because the thermal mass of order $\sqrt{\l_0} \, T$, acting as 
an infrared cutoff, generates inverse powers of the coupling constant in the
loop integrals. Parwani \cite{parwani} showed the consistency
of the HTL resummation up to two--loop order for the case of $\l \phi^4$ 
theory. In the following, we will refer to his results generally as ``the
HTL approximation''.

Up to one loop in the present scalar theory, the HTL resummation amounts
merely to a resummation of daisy diagrams, which systematically incorporates
the thermal mass to order $\sqrt{\l_0} \, T$ into the bare propagator. However,
in the UHT limit the HTL approximation does not properly take the thermal 
corrections to the coupling constant into account. These thermal fluctuations 
are crucial for a correct description of the phase transition. More generally,
we will compare our expression for the coupling constant,
\be
\l (\tau) = \frac{\ds 1}{\ds \frac{1}{\l_0} + \frac{3}{2} \, 
\loopd2^{\!\!\prime} (z(\tau))} \:, \label{ourlambda}
\ee
with the HTL one--loop result \cite{parwani}
\be
\l_\mathrm{HTL} (\tau) = \l_0 - \frac{3}{2} \l_0^2 \, \loopd2^{\!\!\prime} 
(z(\tau)) \:. \label{HTLlambda}
\ee
In Eq.\ (\ref{ourlambda}), we have neglected the logarithmic term which 
becomes notable only at extremely high temperatures, consequently
$\loopd2$ is replaced by $\loopd2^{\!\!\prime}$, defined as the negative of
the second $M^2$--derivative of $\loopd{}^{\,\,\prime}$ where the first,
explicitly $\tau$--independent term is left out in Eq.\ 
(\ref{loopdef}). In the original one--loop HTL expansion, the mass 
appearing in the loop integral in Eq.\ (\ref{HTLlambda}) is replaced by 
its expansion to first order in $\sqrt{\l_0}$, a 
difference which we neglect here for the sake of easier comparison. We also 
include the contribution of the zero--temperature mass $M_0$ in $z (\tau) = 
M(\tau)/\tau$, which is usually put to zero in the HTL approach.
We consider the discrepancy between the two formulas (\ref{ourlambda}) and
(\ref{HTLlambda}) as appreciable as soon as
\be
\frac{\l (\tau) - \l_\mathrm{HTL} (\tau)}{\l_\mathrm{HTL} (\tau)} \approx 
\nu_0 \:, \label{HTLcrit}
\ee
where we typically will take $\nu_0=0.1$, corresponding to a 10\% difference 
between the two values, which is equivalent to
\be
\frac{\ds \frac{3}{2} \, \loopd2^{\!\!\prime} (z)}{\ds \frac{1}{\l_0}} 
\approx 0.3 \:. \label{HTLcrit2}
\ee
The latter relation defines a value of $z$ for any given $\l_0$, which is
the value represented by the HTL curves in the figures below. 

\begin{figure}
\begin{center}
\pspicture(10,6.3)
\rput[lb]{0}(0,-0.5){\includegraphics[width=10cm]{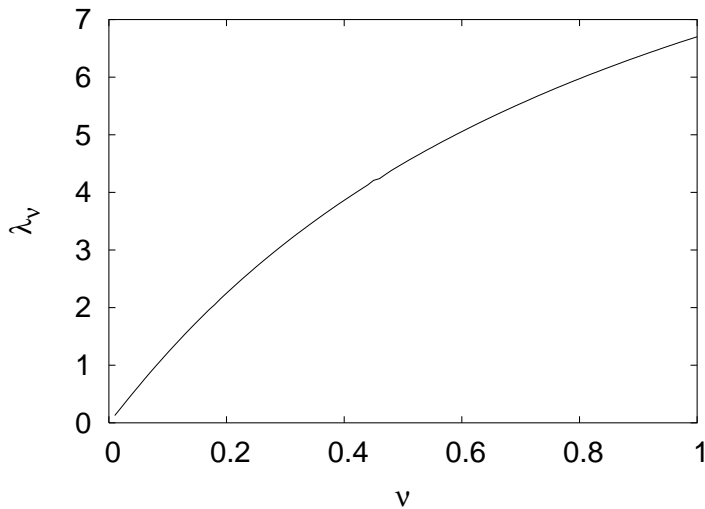}}
\endpspicture
\end{center}
\caption{ The HTL--critical coupling constant $\l_\nu$ as a function of
the parameter $\nu$ specifying the criterion according to Eq.\ 
(\ref{nucrit}). \label{lanu}}
\end{figure}

Our criterion (\ref{HTLcrit}) for the validity of the HTL approximation
might be criticized as subjective. To improve on this matter, observe that
there is precisely one value of the initial coupling constant $\l_0$
such that the equation (\ref{HTLcrit}) is exactly fulfilled in the limit
of large $\tau$ (still neglecting any logarithmic contribution). We call
this value the HTL--critical coupling. Generalizing (\ref{HTLcrit}) to
\be
\lim_{\tau \to \infty} \frac{\l (\tau) - \l_\mathrm{HTL} (\tau)}
{\l_\mathrm{HTL} (\tau)} = \nu \:, \label{nucrit}
\ee
we can define an HTL--critical coupling $\l_\nu$ for every value of $\nu$.
For all initial coupling constants $\l_0 < \l_\nu$, the HTL approximation 
will then be acceptable according to the $\nu$--criterion, for 
sufficiently high temperatures. In Fig.\ \ref{lanu}, $\l_\nu$ is plotted as a
function of $\nu$, for $0 \le \nu \le 1$. 

\begin{figure}
\begin{center}
\pspicture(16,5)
\rput[lb]{0}(0,-0.5){\includegraphics[width=8cm]{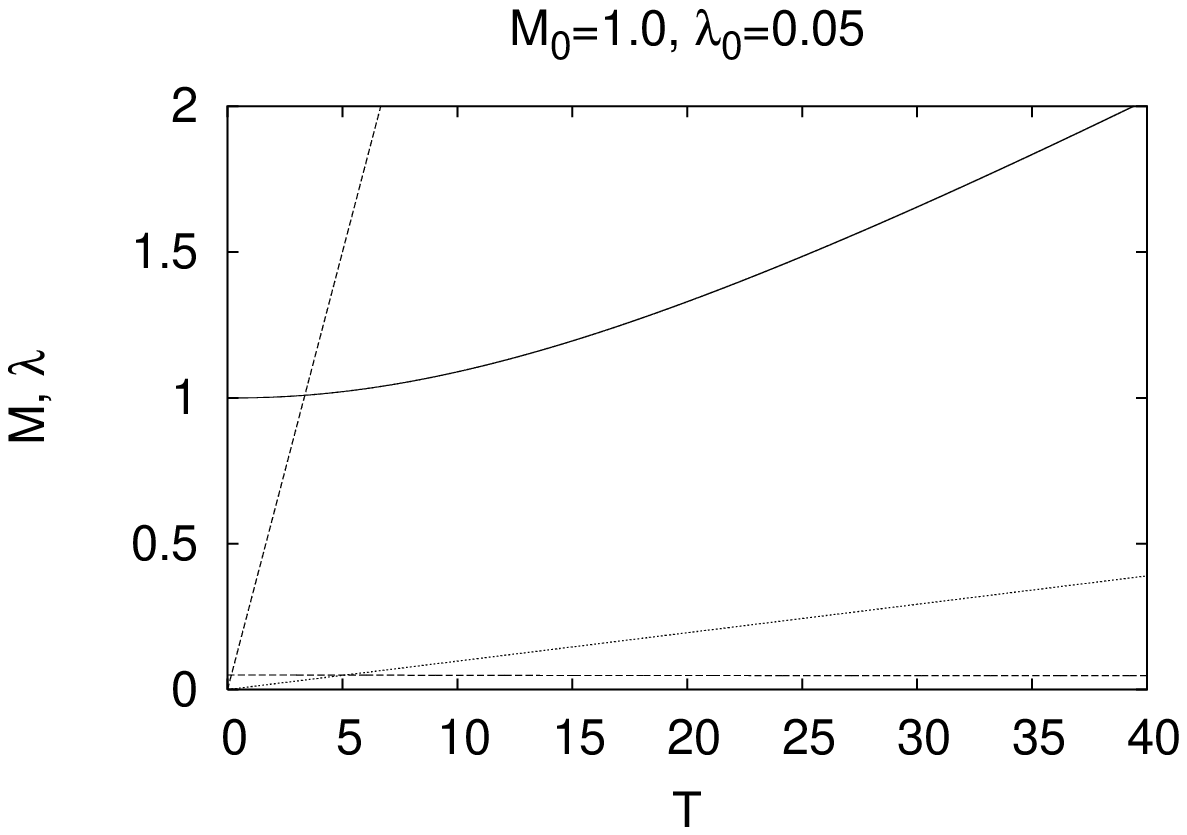}}
\rput[lb]{0}(8,-0.5){\includegraphics[width=8cm]{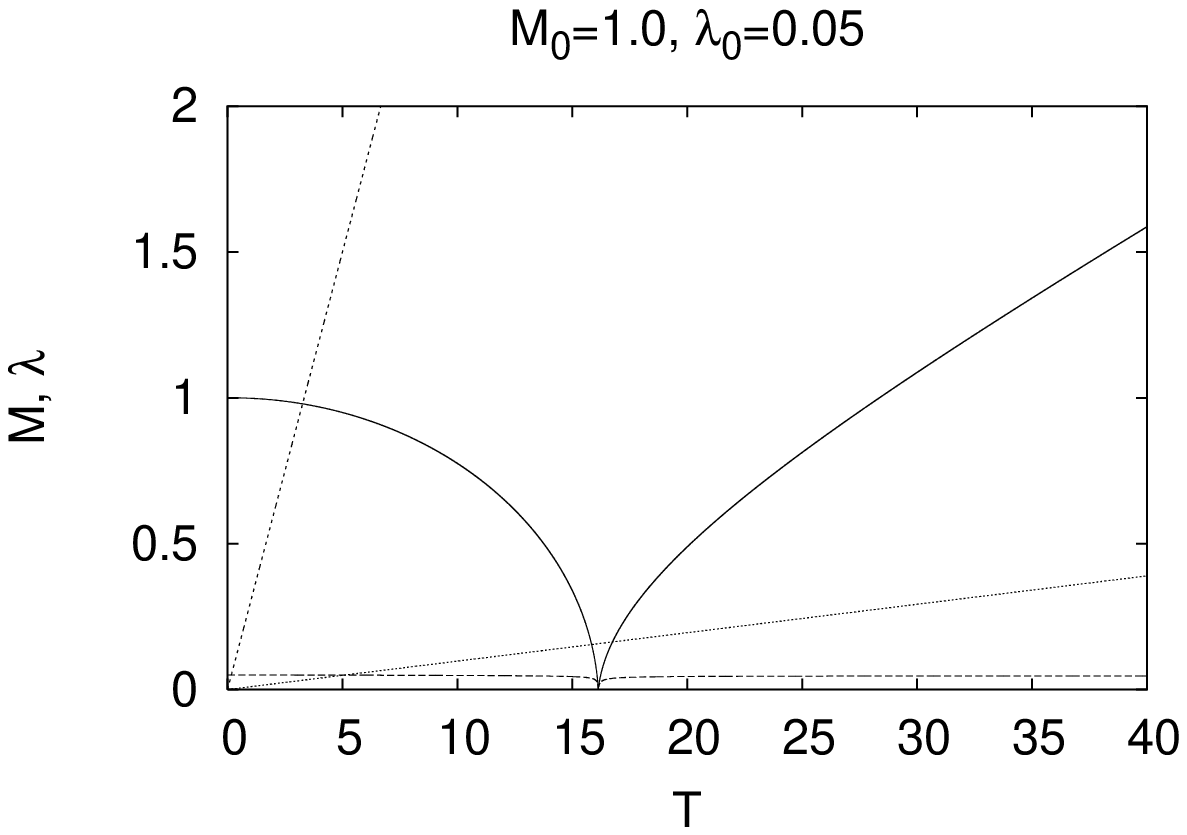}}
\endpspicture
\end{center}
\caption{ Mass $M$ (solid line) and coupling constant $\lambda$ (dashed
line) as a function of the temperature $T$, as obtained from integrating 
differential equations (\ref{massflow}--\ref{couplingflow}) 
with the initial conditions
$M_0 \equiv M (T=0) = 1.0$ (in arbitrary units) and $\lambda_0 \equiv
\lambda (T=0) = 0.05$. For the plot to the left, the system is in the
symmetric phase at $T=0$, while for the plot to the right, it
is in the phase of broken symmetry. The lines of constant $z(T) = M(T)/T$
correspond to the criteria for dimensional reduction (short dashed line) and
HTL approximation (dotted line) discussed in the text. \label{l005m1}}
\end{figure}

\begin{figure}
\begin{center}
\pspicture(16,5)
\rput[lb]{0}(0,-0.5){\includegraphics[width=8cm]{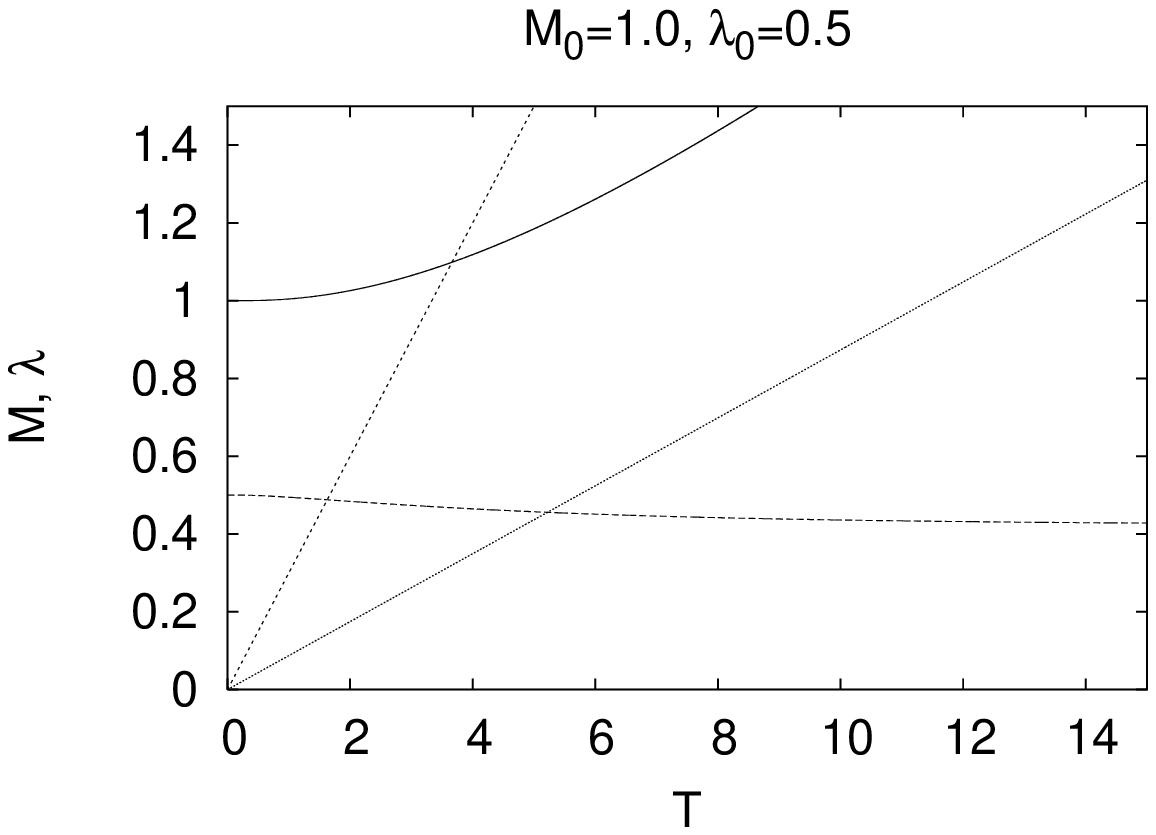}}
\rput[lb]{0}(8,-0.5){\includegraphics[width=8cm]{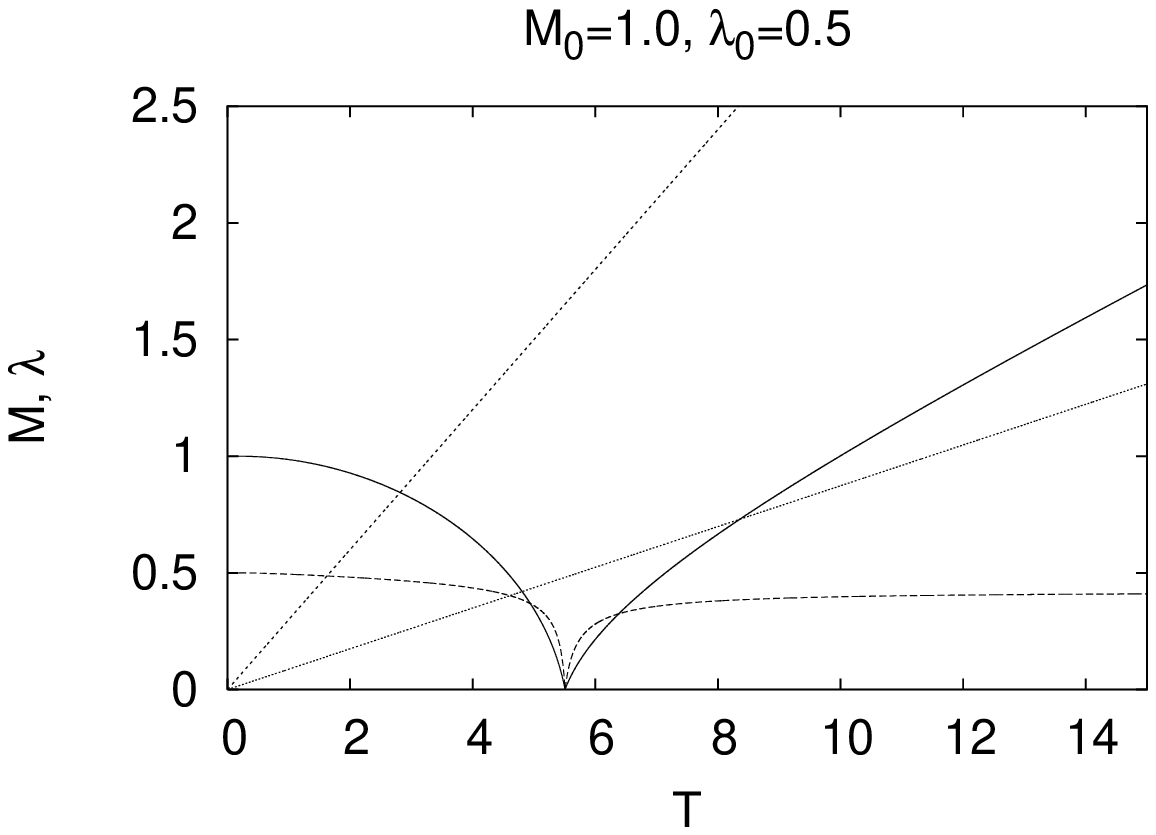}}
\endpspicture
\end{center}
\caption{ As in Fig.\ \ref{l005m1}, but with initial condition $\lambda_0 =
0.5$. \label{l05m1}}
\end{figure}

\begin{figure}
\begin{center}
\pspicture(16,5)
\rput[lb]{0}(0,-0.5){\includegraphics[width=8cm]{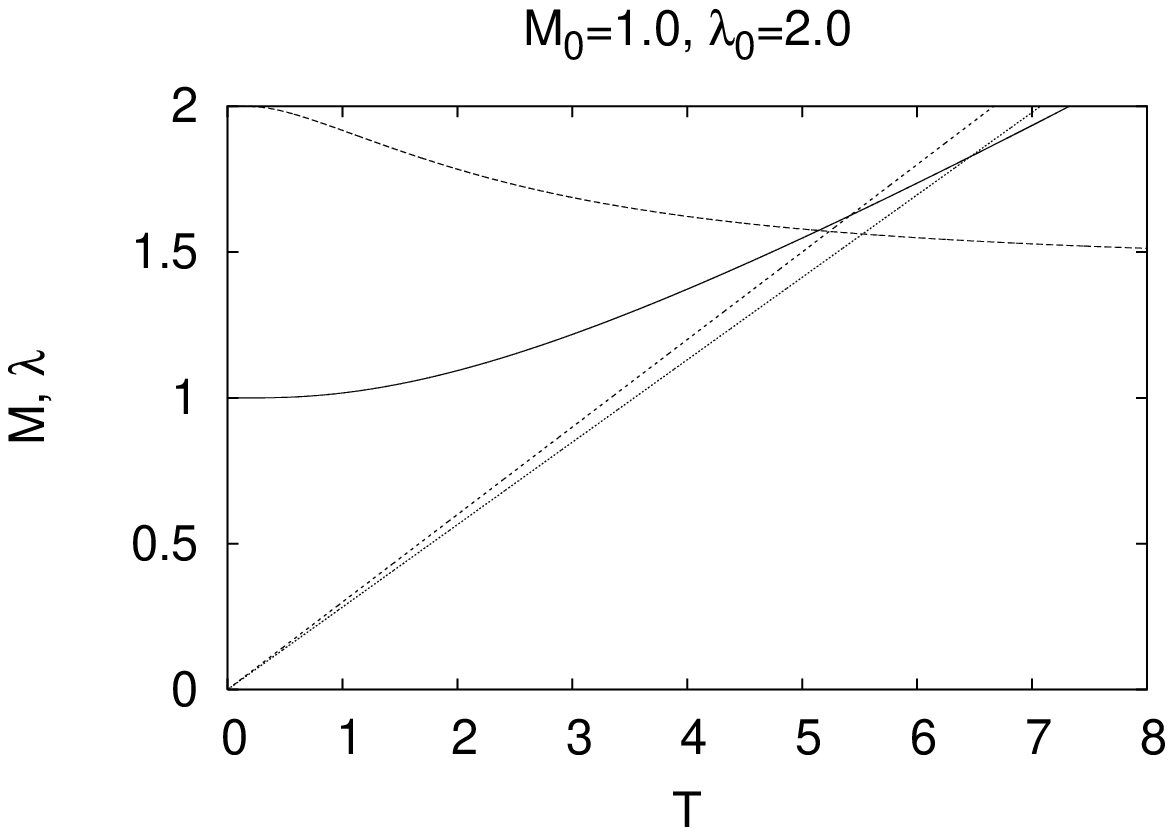}}
\rput[lb]{0}(8,-0.5){\includegraphics[width=8cm]{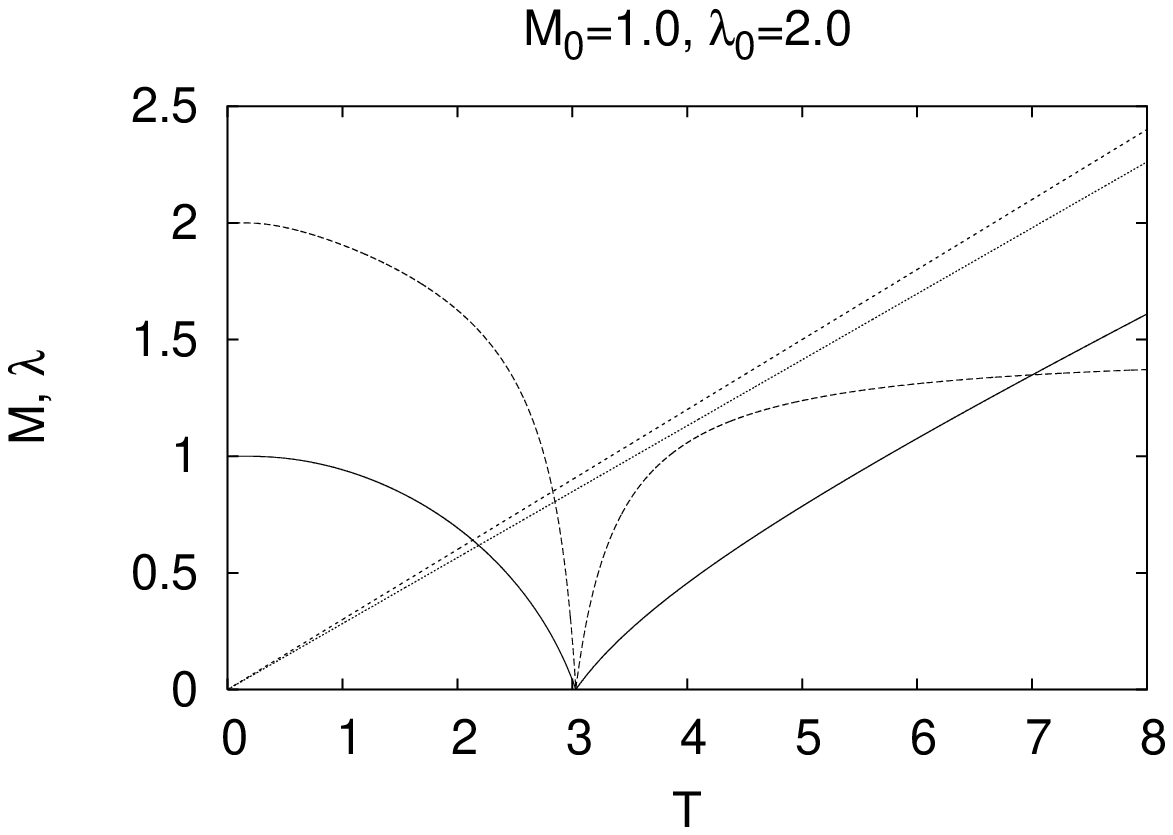}}
\endpspicture
\end{center}
\caption{ As in Fig.\ \ref{l005m1}, but with initial condition $\lambda_0 =
2.0$. \label{l2m1}}
\end{figure}

Having motivated our criteria for dimensional reduction and the validity
of the HTL expansion we will now consider the explicit results: 
In all the figures, the temperatures where dimensional reduction
(approximately) applies are the ones where the curve $M(T)$ lies
{\em below}\/ the dimensional reduction curve (short dashed line), 
whereas the HTL resummation is adequate wherever $M(T)$ lies {\em above}\/ 
the HTL curve (dotted line). In Figs.\ \ref{l005m1}--\ref{l2m1} to the 
left, 
we see graphs of $M(T)$ and 
$\lambda(T)$ as a function of $T$ for the three different initial 
couplings $0.05$, $0.5$ and $2.0$ respectively where we start at $T=0$ in 
the symmetric phase with $M_0=1$. For small coupling, $\lambda_0=0.05$, 
we see that the HTL approximation is always valid and that there is a 
non--universal dimensional reduction starting at about $T=3M_0$. 
Note that straightforward perturbation theory  is expected to
be applicable as long as $\sqrt{\l_0}\, T \ll M_0$, hence in the present
case for $T \ll 4 M_0$. For larger couplings $\lambda_0=2$ NHT
dimensional reduction does not set in until about $T=5M_0$ and the HTL
approximation is only valid for $T<6M_0$. One immediate conclusion of this
is that the existence and the value of the ``high'' temperature above 
which HTL is invalid is sensitively dependent on the initial coupling. 
Hence, contrary to current folklore, even for large temperatures, dimensional 
reduction and HTL {\em only}\/ apply for sufficiently small initial values 
$\l_0$, from Figs.\ \ref{l05m1}, \ref{l2m1} roughly $\l_0 < 2.0$ and 
$\l_0 < 1.0$, respectively.
The breakdown of the HTL approximation is due to the fact that
thermal corrections to the coupling are significant, as at high temperatures
they will vary as $\l_0 \sqrt\l_0$ which when resummed gives rise to a
large contribution.

Note that if we are more rigorous about our 
criterion for validity of the HTL expansion, for example demanding agreement
with the full thermal coupling to 5\% rather than 10\%, then the HTL line
will be steeper and the critical coupling for which a full fluctuation 
analysis is required in the high--temperature limit correspondingly smaller. 
Also, if we require a stricter criterion for dimensional
reduction then the corresponding dimensional reduction line
would be flatter. Clearly, for a certain value of the 
zero--temperature coupling constant depending on the criteria chosen,  
the HTL and dimensional reduction lines cross thus indicating that the regime
of dimensional reduction can no longer be described within the HTL 
approximation if the coupling constant is larger than this critical
value.

In Figs.\ \ref{l005m1}--\ref{l2m1} to the right, we see analogous results to 
those in the same figures to the left, but now for the case where we start 
in the broken symmetry phase with $M_0=1$. In this case there exists a 
phase transition at a critical temperature $T_c$. At $T_c$, the mass as well 
as the effective coupling constant (when defined as in (\ref{normcnds})) 
vanish. For
large temperatures, the mass grows proportionally to the temperature and
the coupling constant tends towards a constant value. For both quantities,
the values at large temperatures become independent of the $T = 0$ initial
condition $M_0$, while they do vary with $\l_0$. Notice in particular that
the asymptotic value of $\l$ can deviate appreciably from $\l_0$, at
least if $\l_0$ is sufficiently large. In actual fact, at 
temperatures very much larger than the ones shown in Figs.\ 
\ref{l005m1}--\ref{l2m1}, $M(T)/T$ and $\l(T)$ show a logarithmic rise 
which originates in the usual renormalization of the UV divergence in the 
one--loop contribution to $\G_t^{(4)}$, and finally leads into a Landau
pole. The high--temperature behaviour of the solutions will be analyzed in 
more detail below. In Figure \ref{l005m1}, 
for $\lambda_0=0.05$, we see that the HTL approximation
is valid except in a small region around the critical temperature, of width 
about $0.1T_c$, and that there is a non--universal dimensional reduction 
starting at about $T=0.2T_c$ with a universal dimensional reduction setting
in much closer to the critical temperature where universal quantities are 
essentially three--dimensional. On the other hand, in Figure \ref{l2m1}, 
for $\lambda_0=2.0$, we see that non--universal dimensional reduction
does not set in until $T>0.7T_c$ and moreover that the HTL approximation
breaks down at about the same temperature. Once again, the critical value
of the coupling constant where the HTL and the dimensional reduction lines
cross and hence the regime of dimensional reduction cannot be described 
within the HTL approximation, depends on the criteria chosen. The important 
conclusion to be drawn here is that there may be no non--universal 
dimensional reduction via which an effective three--dimensional theory may 
be posited before universal fluctuations invalidate the use of an HTL 
approximation necessary to analytically calculate the parameters of
the three dimensional theory as is done in 
the context of the electroweak theory in \cite{shaposhnikov}. Once again 
we emphasize that our present estimates are based on a liberal 
criterion for granting dimensional reduction and a fairly liberal one for 
determining the validity of the HTL approximation.  

Several properties of the solutions of the flow equations 
(\ref{massflow}--\ref{couplingflow}), and in particular the limit of high 
temperatures, become more transparent when one considers the flow in the 
$z$--$\l$--plane.
To begin with, we will neglect the logarithmic contributions which become
important only at extremely high temperatures, thus considerably simplifying 
the following discussion. The integration of the flow equation for $\l$
defines, after neglecting the logarithmic term, a characteristic curve 
$\l(z)$ along which a fixed physical system evolves with changing 
temperature. By considering the limit $\tau \to 0$ or $z \to \infty$ we infer 
that the characteristic curve is specified by {\em only}\/ choosing the 
initial coupling constant $\l_0$. Furthermore, the matching condition 
(\ref{match}) implies that the curve is the same below and above the
critical temperature, for a fixed physical system (which is in the
broken--symmetry phase at $\tau = 0$).

We now determine the flow of the system along its characteristic curve
for increasing temperature. The flow equations for the mass imply that
\be
\tau \frac{d z}{d \tau} = - \frac{\l}{2 M} \frac{\p}{\p \tau} 
\loopd1 - \frac{3 \l M}{4} \frac{\p}{\p \tau} \loopd2 - z \label{ztaubroke}
\ee
in the phase of broken symmetry, and
\be
\tau \frac{d z}{d \tau} = \frac{\l}{4 M} \frac{\p}{\p \tau} \loopd1 - z
\label{ztausymm}
\ee
in the symmetric phase. The right--hand sides of these equations turn out
to depend only on $z$ and $\l$, hence the change of $z$ with $\tau$ (and
consequently the change of $\l$ with $\tau$) is uniquely determined for
every point on the characteristic curve in the $z$--$\l$--plane. As for
the direction of the flow, the right--hand side of Eq.\ (\ref{ztaubroke})
is manifestly negative, hence in the broken--symmetry phase the system flows 
from $z \to \infty$ (at $\tau = 0$) towards smaller $z$ until it arrives,
at $\tau = T_c$, at $z = 0$.

The right--hand side of Eq.\ (\ref{ztausymm}), however, can have both
signs, depending on the value of $z$ ($\l$ is a well--defined function of 
$z$ for a fixed characteristic curve). It is positive for small $z$ and
negative for large $z$, hence the flow with growing temperature is towards
an intermediate (quasi--) fixed point determined by
\be
z = \frac{\l (z)}{4 M} \frac{\p}{\p \tau} \loopd1 \:, \label{fixp}
\ee
which defines the asymptotic high--temperature values of $z$ and $\l$.
For systems which enter the symmetric phase after a phase transition at
finite temperature, the flow in the symmetric phase starts at $z = 0$
(corresponding to $\tau = T_c$) and tends towards larger values of $z$.
For systems which are in the symmetric phase at $\tau = 0$, the flow
starts at $z \to \infty$ and is directed towards smaller values of $z$.
In both cases, for $\tau \to \infty$, the fixed point (\ref{fixp}) is
reached asymptotically.

\begin{figure}
\begin{center}
\pspicture(10,6.3)
\rput[lb]{0}(0,-0.5){\includegraphics[width=10cm]{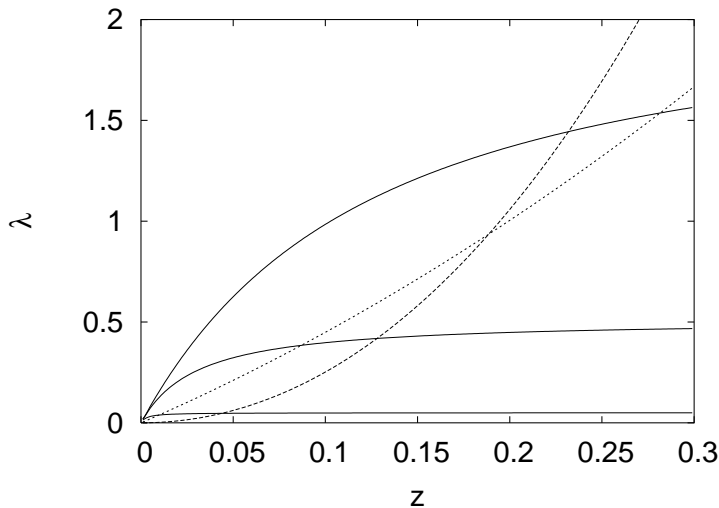}}
\endpspicture
\end{center}
\caption{ The characteristic curves in the $z$--$\lambda$--plane for
initial values $\lambda_0 = 0.05$, $0.5$, $2.0$ (solid curves), together with
the line of quasi--fixed points (dashed line) and the HTL curve (dotted line)
discussed in the text. \label{la2}}
\end{figure}

The characteristic curves for the initial values $\l_0 = 0.05, 0.5, 2.0$
are plotted in Fig.\ \ref{la2}, together with the line of
(quasi--) fixed points (dashed).
 The line for the validity of the HTL approximation is also represented 
in Fig.\ \ref{la2} (dotted). It intersects every characteristic curve once,
and at the intersection Eq.\ (\ref{HTLcrit2}) is fulfilled. As for the
curves corresponding to dimensional reduction in Figs.\ 
\ref{l005m1}--\ref{l2m1}, they simply correspond to vertical lines at 
$z = z_0$ ($z_0=0.3$ throughout most of the paper). The HTL approximation
is adequate for the part of the characteristic curve to the right of the
HTL line (far from the phase transition where the fluctuations in $\l$
are important), while dimensional reduction applies to the left of
$z = 0.3$ (close to the phase transition).

For small $z$, all loop integrals can be evaluated analytically 
\cite{vaneijck}. One then has that all characteristic curves are given by
\be
\l (z) = \frac{16 \pi}{3} \, z + {\cal O} (z^2 \ln z) \:,
\ee
independently of $\l_0$. The universal slope is a reflection of universality
at the critical point and originates exclusively from the thermal
fluctuations. The HTL curve is given by
\be
\l (z) = \frac{16 \pi}{3} \, \frac{\ds z}{\ds 1 + \sqrt{\frac{1 + \nu}{\nu}}}
+ {\cal O} (z^2 \ln z) \:,
\ee
where we have considered a general $\nu$ instead of $\nu_0 = 0.1$ in Eq.\ 
(\ref{HTLcrit}). It is then clear that for sufficiently small $z$, all
characteristic curves lie above the HTL line (for {\em any}\/ choice of
$\nu$), so that the HTL approximation does not apply. Finally, for the line 
of high--temperature fixed points, we have
\be
\l (z) = 24 z^2 + {\cal O} (z^3) \:,
\ee
so that the fixed point line lies below the HTL curve (again for any
choice of $\nu$) for sufficiently small $\l (z)$. As a consequence, at high
temperatures the HTL approximation is satisfactory as long as $\l_0$ 
(and hence $\l (\tau)$ for $\tau \to \infty$) is sufficiently small.

The representation of the flow in the $z$--$\l$--plane makes the
high--temperature limit particularly transparent through the line of
(quasi--) fixed points. First of all, it is clear that there are really
asymptotic values for $\l (\tau)$ and for $M(\tau)/\tau$. Furthermore,
these values {\em only}\/ depend on $\l_0$ and not on the zero temperature
mass $M_0$, nor on the phase the system is in at $\tau = 0$. The 
independence of the physics at high temperatures on $M_0$ is part of the 
usual folklore. However, it is also apparent in this representation 
that dimensional reduction
and the HTL approximation do not automatically apply at high enough
temperatures, but that they require the smallness of the coupling constant.
Finally, the asymptotic value $\l (\tau)$ for $\tau \to \infty$ is always
smaller than $\l_0$ and can be calculated for a given $\l_0$ using Eq.\
(\ref{fixp}) and the integrated flow equation for $\l$. In an expansion
in $\l_0$ around $\l_0 = 0$, one has
\be
\l_\infty \equiv \lim_{\tau \to \infty} \l (\tau) = \l_0 \left[ 
1 - \frac{9}{2 \pi} \sqrt{\frac{\l_0}{24}} \right] + {\cal O}(\l_0^2
\ln \l_0) \:. \label{lambdapert}
\ee
Analogously, one can determine the asymptotic value for $z (\tau)$,
$\tau \to \infty$, and obtains for the mass in a small--$\l_0$ expansion
\be
\lim_{\tau \to \infty} M^2 (\tau) = \frac{\l_\infty \tau^2}{24} \left[ 1 - 
\frac{3}{2 \pi} \sqrt{\frac{\l_\infty}{24}} \right] + {\cal O}(\l_\infty^2)
= \frac{\l_0 \tau^2}{24} \left[ 1 - \frac{6}{\pi} \sqrt{\frac{\l_0}{24}} 
\right] + {\cal O}(\l_0^2 \ln \l_0) \:. \label{masspert}
\ee

In the HTL approach, Eq.\ (\ref{lambdapert}) is equally valid, which is
a direct consequence of the fact that Eqs.\ (\ref{ourlambda}) and
(\ref{HTLlambda}) coincide to the order considered here. However, and
somewhat surprisingly, Parwani \cite{parwani} finds for the
thermal mass
\be
\lim_{\tau \to \infty} M_\mathrm{HTL}^2 (\tau) = \frac{\l_0 \tau^2}{24} 
\left[ 1 - \frac{3}{\pi} \sqrt{\frac{\l_0}{24}} 
\right] + {\cal O}(\l_0^2 \ln \l_0) \:,
\ee
which is clearly different from our result (\ref{masspert}). In order to
decide which of the two resummation schemes is more appropriate in the
sense of a faster convergence towards the exact result, one would have
to calculate the higher--order contributions in both schemes. We will
not do that here, but rather consider the analogous problem in the limit 
of large $N$ in the $O(N)$ model, which will shed at least some light on 
this issue.

In the $O(N)$ model in the {\em symmetric}\/ phase, the flow
equations take a form nearly identical with Eqs.\ (\ref{massflow}, 
\ref{couplingflow}), one only has to replace on the right--hand sides
\be
\loopd1 \longrightarrow \frac{N + 2}{3} \, \loopd1 \:, \quad
\loopd2 \longrightarrow \frac{N + 8}{9} \, \loopd2 \:. \label{subsN}
\ee
Repeating the steps leading to Eq.\ (\ref{masspert}) gives the following
result for the expansion in $\l_0$ of the thermal mass at high temperatures,
\be
\lim_{\tau \to \infty} M^2 (\tau) = \frac{N + 2}{3} \,
\frac{\l_0 \tau^2}{24} \left[ 1 - \left( \frac{N + 5}{N + 2} \right) 
\frac{3}{\pi} \sqrt{\frac{N + 2}{3} \, \frac{\l_0}{24}} \right] + 
{\cal O}(\l_0^2 \ln \l_0) \:. \label{Nmass}
\ee
For $N=1$, we of course recover the result (\ref{masspert}). Now,
performing the analogue of the HTL analysis for the $O(N)$ model, one
arrives at
\be
\lim_{\tau \to \infty} M_\mathrm{HTL}^2 (\tau) = \frac{N + 2}{3} \,
\frac{\l_0 \tau^2}{24} \left[ 1 - \frac{3}{\pi} 
\sqrt{\frac{N + 2}{3} \, \frac{\l_0}{24}} \right] + 
{\cal O}(\l_0^2 \ln \l_0) \:. \label{NHTL}
\ee
These new results {\em only}\/ coincide in the limit $N \to \infty$. This
is a particularly interesting limit insofar as the theory is exactly
solvable for $N \to \infty$. The limit has to be taken in such a way that
\be
\bar{\l}_0 \equiv \frac{N + 2}{3} \, \l_0 \label{largeNl}
\ee
tend towards a (non--vanishing) constant. As we shall soon see, the 
expression found above for $\lim_{\tau \to \infty} M^2 (\tau)$ in the
$N \to \infty$ limit is correct (to this order in $\l_0$).

As we mentioned above, the $O(N)$ model can be exactly solved in the
large--$N$ limit. The result for $M(\tau)$ and $\bar{\l}(\tau)$ defined
as in Eqs.\ (\ref{massCnd}, \ref{normcnds}) and (\ref{largeNl}), is in the 
{\em symmetric}\/ phase
\be
M^2(\tau) &=& M_0^2 + \frac{\bar{\l}_0}{2} \loopd1^{\!\!\prime} 
(M(\tau), \tau) \:, \label{superdaisy} \\
\bar{\l} (\tau) &=& \frac{\ds 1}{\ds \frac{1}{\bar{\l}_0} + \frac{1}{2}
\loopd2^{\!\!\prime} (M(\tau), \tau)} \:, \label{bubble}
\ee
where we still have neglected any logarithmic term. We can describe a
phase transition at a finite temperature by choosing the parameter $M_0^2$
negative (but the above equations are only valid above the critical
temperature). Eq.\ (\ref{superdaisy})
is an implicit equation for $M(\tau)$. An iterative solution leads
to the set of the so--called superdaisy diagrams \cite{dolanjackiw}. 
Eqs.\ (\ref{Nmass}) (for $N \to \infty$) and (\ref{NHTL}) are the 
result of the second iteration in the limit of high temperatures
and correspond to the daisy diagrams. Eq.\ (\ref{bubble}), when expanded in a 
geometric series, generates the complete series of bubble or chain diagrams 
where every bubble in itself corresponds to the sum of all superdaisy 
diagrams (with two external vertices).

It is then easy to verify that the total $\tau$--derivative of Eqs.\
(\ref{superdaisy}) and (\ref{bubble}) leads exactly to the flow equations
(\ref{massflow}, \ref{couplingflow}) with the substitutions (\ref{subsN}) 
in the limit
of large $N$. In other words, the integration of the one--loop
flow equations leads to the {\em exact}\/ solution in the large--$N$ limit.
This fact gives a lot of confidence in the results of the integration of
the flow equations also for finite $N$, in particular, in Eq.\
(\ref{masspert}) for $N=1$. Also observe that in the HTL formalism, the
{\em complete}\/ (resummed) perturbative series has to be summed up to
obtain the results (\ref{superdaisy}) and (\ref{bubble}). As a last
remark, the use of the total instead of the partial $\tau$--derivative in 
the flow equation (\ref{massflow}) for the mass, which might have been 
naively expected, can now clearly be seen to be inconsistent,
since such flow equations would lead to a result different from
(\ref{superdaisy}) and (\ref{bubble}) in the large--$N$ limit, through
overcounting of diagrams.

In large parts of the foregoing discussion, we have neglected the
logarithmic contributions with the argument that they are negligible for
all but extremely high temperatures. This approximation, apart from
bringing out properties of the flow equations that would otherwise hardly
be visible in an analytic approach, also simplifies the comparison with
the HTL approximation and the superdaisy diagram summation {\`a} la
Dolan--Jackiw \cite{dolanjackiw}. For the rest of this section, 
we will discuss the effects of including the logarithms.

Beginning with the large--$N$ limit, we reintroduce the logarithms in
Eqs.\ (\ref{superdaisy}) and (\ref{bubble}) simply by omitting the primes
on the loop integrals. These integrals are then divergent, and we
renormalize by subtracting the pole terms as in a naive
MS renormalization scheme with temperature--dependent counterterms (we
may also subtract the corresponding $\ln (4 \pi) - \gamma_E$, as in the
$\overline{\mathrm{MS}}$ scheme). The result is
\be
M^2(\tau) &=& M_r^2 + \frac{\bar{\l}_r}{2} \left[ \loopd1^{\!\!\prime} 
(M(\tau), \tau) + \frac{M^2 (\tau)}{(4 \pi)^2} \left( \ln \frac{M^2 (\tau)}
{\mu^2} - 1 \right) \right] \:, \label{superdaisylog} \\
\bar{\l} (\tau) &=& \frac{\ds 1}{\ds \frac{1}{\bar{\l}_r} + \frac{1}{2}
\left[ \loopd2^{\!\!\prime} (M(\tau), \tau) - \frac{1}{(4 \pi)^2} \ln
\frac{M^2 (\tau)}{\mu^2} \right]} \:, \label{bubblelog}
\ee
where $M_r^2$ and $\bar{\l}_r$ are the renormalized parameters determined
by the choice of initial conditions $M_0^2$ and $\bar{\l}_0$, 
and $\mu$ sets a (arbitrary) scale for $M(\tau)$.
Taking again the total $\tau$--derivatives of these equations, one recovers 
the flow equations as before, but now including all logarithmic terms. 

\begin{figure}
\begin{center}
\pspicture(16,5)
\rput[lb]{0}(0,-0.5){\includegraphics[width=8cm]{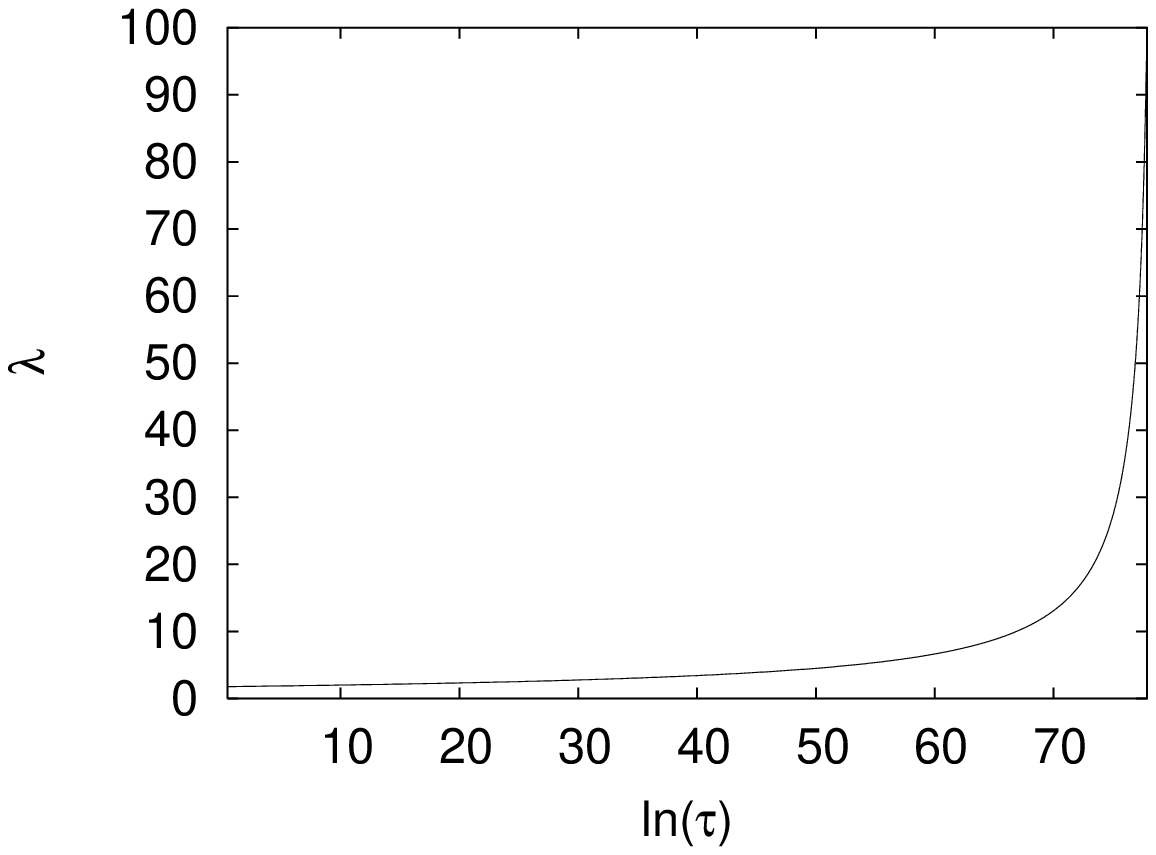}}
\rput[lb]{0}(8,-0.5){\includegraphics[width=8cm]{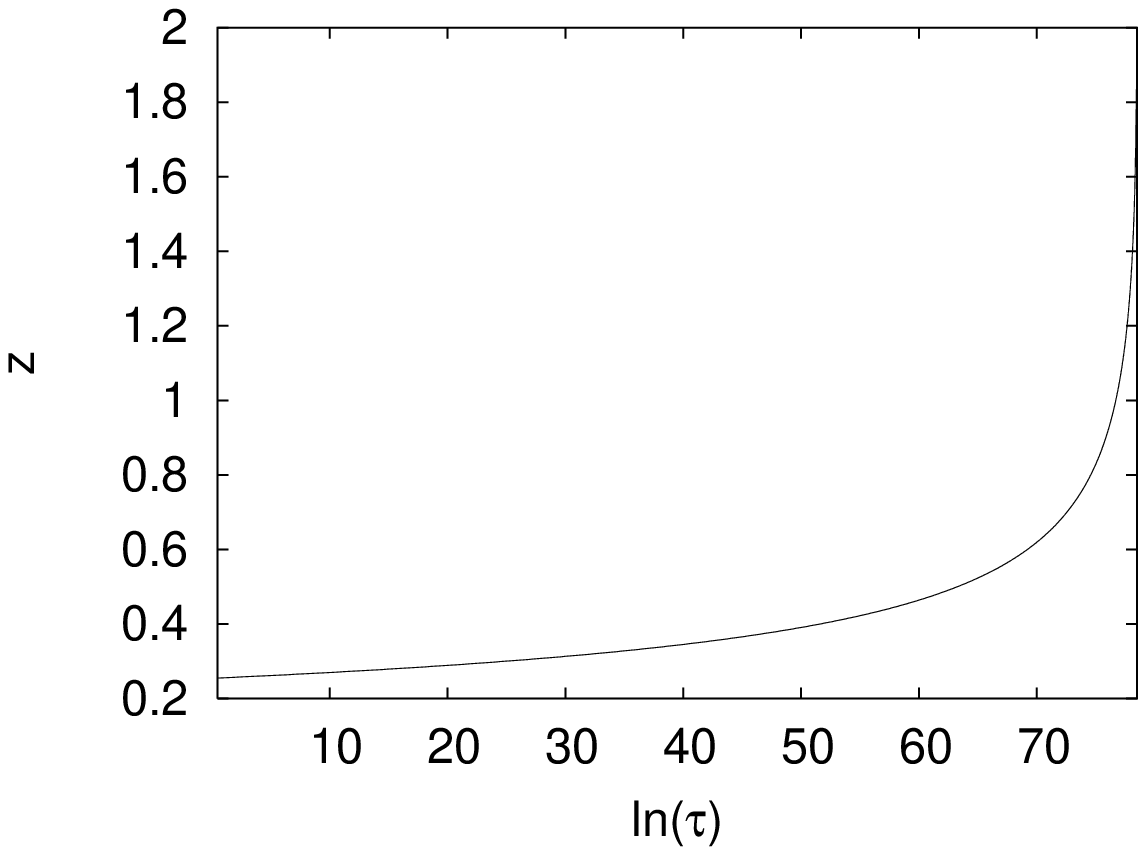}}
\endpspicture
\end{center}
\caption{ Solutions of Eqs.\ (\ref{superdaisylog}, \ref{bubblelog}) for
the temperature dependence of the coupling constant $\bar{\lambda}$ and
$z(\tau) = M(\tau)/\tau$ in the limit of large $N$. At temperature
$\tau = 0$, the system is in the symmetric phase with the initial conditions
$\bar{\lambda}_0 = 2.0$ and $M_0 = \mu$. From Eqs.\ (\ref{superdaisylog}, 
\ref{bubblelog}) one then finds that $\bar{\lambda}_r = \bar{\lambda}_0
= 2.0$ and $M_r^2 = \left[ 1 + 1/(4 \pi)^2 \right] \mu^2$. However, for the
plot we have neglected $M_r^2$ in Eq.\ (\ref{superdaisylog}) since it does 
not contribute in the limit of high temperature which is our primary concern 
here. \label{lam2}}
\end{figure}

In practice, the difference between Eqs.\ (\ref{superdaisylog},
\ref{bubblelog}) and Eqs.\ (\ref{superdaisy}, \ref{bubble}) is only
visible for very large temperatures. We present the solution of Eqs.\
(\ref{superdaisylog}, \ref{bubblelog}) for the initial condition
$\bar{\lambda}_0 = 2.0$ in Fig.\ \ref{lam2}.
 Notably, the values of $\bar{\lambda}$ and $z$ do not vary appreciably 
over a large range of temperatures and stay close to the quasi--fixed point
determined from Eqs.\ (\ref{superdaisy}, \ref{bubble}), i.e., by neglecting 
the logarithmic terms (for $\bar{\l} = 2.0$, the quasi--fixed point is at
$z = 0.2570$). This is the significance of the quasi--fixed point for the
complete system of equations (with logarithmic terms). Finally, at 
$\ln (\tau/\mu) = 78.47$, $\bar{\lambda}$ diverges while $z$ tends 
towards the finite value 2.05. This behavior
is a reflection of the Landau pole at high momentum in the 
zero--temperature theory.

We mention that the logarithmic terms originating from the correct
renormalization of the corresponding diagrams, also appear in the original 
work of Dolan and Jackiw \cite{dolanjackiw} and in the HTL expansion
\cite{parwani}. However, since the latter expansion is organized in powers 
of $\sqrt{\l_0}$ and its logarithm, the logarithms of the one--loop diagrams 
are grouped together with contributions from two--loop diagrams.

\begin{figure}
\begin{center}
\pspicture(16,5)
\rput[lb]{0}(0,-0.5){\includegraphics[width=8cm]{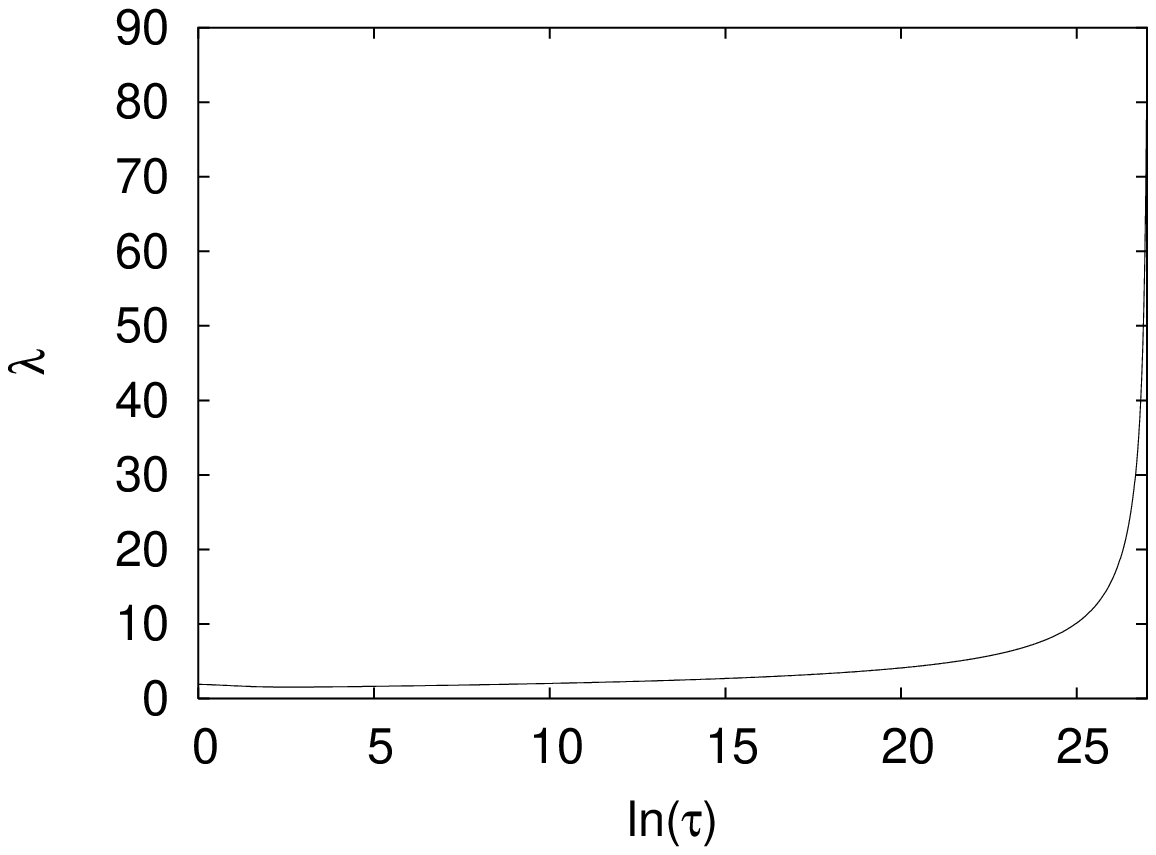}}
\rput[lb]{0}(8,-0.5){\includegraphics[width=8cm]{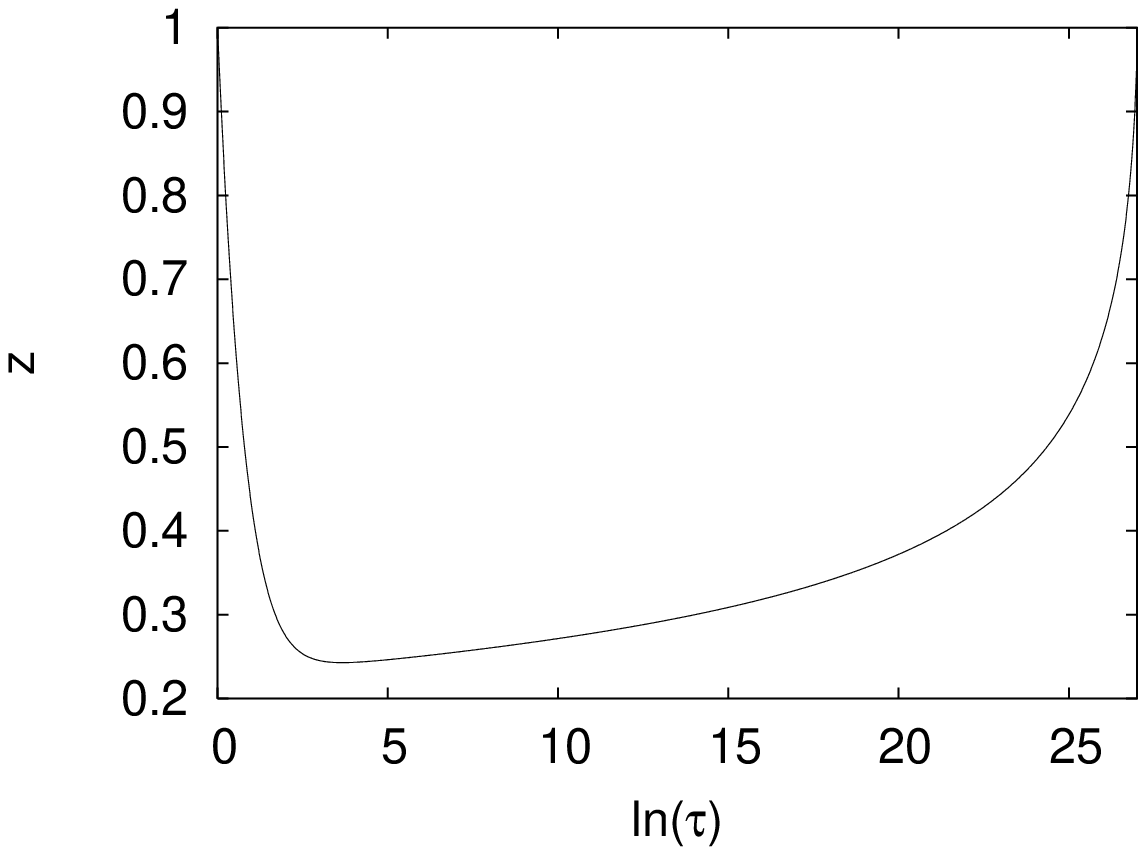}}
\endpspicture
\end{center}
\caption{ Solutions of Eqs.\ (\ref{massflow}, \ref{couplingflow}) for 
the temperature dependence of the coupling constant $\lambda$ and 
$z(\tau) = M(\tau)/\tau$ for $N=1$. At temperature $\tau = 0$, the system is 
in the symmetric phase with the initial conditions $\lambda_0 = 2.0$ 
and $M_0 = 1$ (in arbitrary units). \label{lam2-ln}}
\end{figure}

Turning now to $N=1$, we show the solutions of Eqs.\ (\ref{massflow},
\ref{couplingflow}) over a large range of temperatures in Fig.\ 
\ref{lam2-ln}, again for $\lambda_0 = 2.0$.
Observe that the qualitative behaviour is as in Fig.\ \ref{lam2}, i.e.,
for large $N$ (note that for Fig.\ \ref{lam2} we have neglected $M_r^2$
in Eq.\ (\ref{superdaisylog}) which is why the curves for $z (\tau)$ look
different for small $\tau$). There is a plateau of nearly constant values
of $\lambda$ and $z$ for a large temperature range, close to the quasi--fixed
point of Eq.\ (\ref{fixp}) (neglecting the logarithmic contributions). For
the present case with $\lambda_0 = 2.0$, the quasi--fixed point is at 
$z = 0.232$. For smaller initial couplings, the plateau is considerably more
pronounced --- flatter as well as wider, and closer to the quasi--fixed point 
values. For very large temperatures, eventually $\lambda$ runs into the
analogue of a Landau pole, at a finite value of $z$. We remark that the
integration of the differential equations (\ref{massflow}, 
\ref{couplingflow}) over such
a large range of temperatures is a highly non--trivial task from a numerical
point of view, and our confidence in the results obtained is partly based
on the very similar behaviour of the solutions of the algebraic equations
(\ref{superdaisylog}, \ref{bubblelog}), which we checked for different
values of the initial coupling $\lambda_0$.

On the analytical side, it is easy to see that characteristic curves in the 
$z$--$\l$--plane exist even after including the logarithmic term. 
They are determined by the differential equation
\be
\frac{d \l}{d z} = - \frac{3}{2} \l^2 \frac{d}{d z}
\loopd2^{\!\!\prime} (z) + \frac{3}{(4 \pi)^2} \l^2
\left[ \frac{1}{z} + \frac{1}{\ds \tau \frac{d z}{d \tau}} \right] \:,
\ee
where $\tau (d z/d \tau)$ has to be taken from Eq.\ (\ref{ztaubroke}) or
(\ref{ztausymm}), according to the phase. However, analytic solutions of 
this equation are not feasible. Qualitatively, the solutions
approach the line of former high--temperature fixed points from both 
sides at temperatures of
the order of several $M_0/\sqrt{\l_0}$, and move away from the corresponding 
quasi--fixed point only at very much higher temperatures, following the line
of quasi--fixed points very closely and slowly towards larger values of
$z$ and $\lambda$. They eventually lift off the fixed point line to approach
the ``Landau pole'' in $\lambda$ at finite $z$.

The mere existence of the characteristic curves and the quasi--fixed points
implies that the physics at sufficiently high temperatures becomes, to a very
good approximation, independent of the initial value of the mass $M_0$ and the
phase at zero temperature, and only depends on the value of $\lambda_0$.
The HTL approximation eventually breaks down for any value of $\lambda_0$ 
because it is not capable of reproducing a Landau pole without further 
resummation.

\section{Conclusions}

In this paper we have considered the different regimes in which a dimensional
reduction may occur at finite temperature and compared the validity of 
different computational schemes for accessing these domains. We emphasized
that there are two distinct regimes in which dimensional reduction may occur
at high temperature --- with the temperature--dependent mass much
smaller than and of the same order as the characteristic scale $g T$,
respectively --- pointing out that the latter is non--universal 
(NHT regime) whereas the former is universal (UHT regime), 
being independent of the zero temperature parameters. NHT 
dimensional reduction can be successfully described using the HTL 
approximation as in this case the leading order infrared divergences can be 
resummed leading to a thermal mass that acts as an infrared cutoff for 
fluctuations. The resulting behaviour is then mean--field like in that no 
universal fluctuations lead to strong corrections, as in critical phenomena. 
On the contrary, in the UHT regime HTL--type approximations are invalid and 
other methodologies must be sought. We have shown that environmentally 
friendly RGs allow for a complete consistent analysis of both the NHT and 
UHT regimes. 

The region in which the HTL approximation breaks down
is a window around the critical temperature whose size monotonically increases
as a function of the zero temperature coupling. Interestingly, we found 
that for sufficiently large coupling constants, this ``window'' extends
to arbitrarily high temperatures, so that, contrary to common folklore,
the HTL approximation is not guaranteed to be applicable to this regime.
The reason for the breakdown of the HTL approximation are strong
thermal corrections to the coupling constant which cannot be taken into
account correctly without a further resummation. At extremely high
temperatures, we furthermore found a Landau pole for the coupling constant, 
which cannot be described by the HTL expansion to any finite order.
 
A similar window around the critical temperature indicates the 
realization of dimensional reduction. While for small couplings
dimensional reduction applies for arbitrarily high temperatures, for 
sufficiently large couplings it is only valid over a finite temperature 
range. This failure of dimensional reduction for high temperatures
which seems to go against intuition, has its origin in the linear rise of the
thermal mass with temperature (over a wide range of high temperatures) with
a proportionality constant that grows monotonically with the 
zero--temperature coupling.

One of the most interesting conclusions we can glean from our results is that
whether or not there is a regime of non--universal dimensional reduction
which can be described through a HTL--type analysis depends sensitively on 
the magnitude of the zero--tem\-per\-a\-ture coupling constant.
We saw that for sufficiently strong couplings the HTL window around the
critical temperature described above, where the HTL approximation is 
{\em not} valid, completely contains the dimensional reduction window 
where this reduction obtains. Just how strong these couplings have to be
was seen to depend on how rigorous our criteria were for dimensional 
reduction and validity of the HTL approximation. This is an important
finding in view of methods that work with dimensionally reduced
theories and therefore depend on the existence of a regime where the
four--dimensional parameters can be mapped onto three--dimensional ones
through an analytical, HTL--like procedure. In the particular case of the
electroweak phase transition, the existence of such a regime in the
broken--symmetry phase seems to be guaranteed \cite{shaposhnikov}.
At any rate, our results 
further confirm that an environmentally friendly RG offers a general 
methodology for considering high temperature field theory that does not suffer 
from the defects of other methodologies. 

One other defect of the HTL approximation vis \`a vis environmentally 
friendly RGs is in the case where there are Goldstone bosons in the broken 
phase as these lead to universal infrared divergences that cannot be resummed 
\`a la HTL approximation. Environmentally friendly renormalization in 
distinction is capable of treating the Goldstone bosons, their effect being 
to lead to a fixed point for the coupling constant other than the 
Wilson--Fisher fixed point \cite{oconnorstephensphysrep}.

\end{document}